\documentclass[12pt]{article}

\usepackage[english]{babel}
\usepackage{amsmath}
\usepackage{amssymb}
\usepackage{amsthm,amsfonts}
\usepackage{dsfont}
\usepackage{tensor}
\usepackage[dvipsnames]{xcolor}
\usepackage[mathstyleoff]{breqn}
\usepackage{slashed}
\usepackage{stmaryrd}
\usepackage{enumerate}
\usepackage[normalem]{ulem}
\usepackage{tikz}
\usepackage[mathscr]{euscript}
\usepackage{mathrsfs}
\usepackage[numbers,sort&compress]{natbib}
\usepackage{hyperref}
\usepackage{amsthm}
\usepackage{svrsymbols}
\usepackage{phaistos}
\usepackage{mathtools}
\usepackage{stmaryrd} 
\usepackage{graphicx}

\newcommand{\nn}{\nonumber}

\def\ap{\alpha^{\prime}}
\def\ov{\overline}
\def\un{\underline}

\def\be{\begin{equation}}
\def\ee{\end{equation}}
\def\bea{\begin{eqnarray}}
\def\eea{\end{eqnarray}}

\hypersetup{
    colorlinks=true,
    linkcolor=MidnightBlue,
    citecolor=MidnightBlue,
    urlcolor=MidnightBlue
}

\numberwithin{equation}{section} 

\begin{document}
	
\pagestyle{plain}


\pagestyle{empty}
\begin{center}

\vskip .5cm
		
\LARGE{\LARGE\bf Kosmann derivative and momentum maps from a duality covariant framework}

\vskip 0.3cm

\large{Romina Ballesteros$^{a}$, Eric Lescano$^{b}$ and Jesús A. Rodríguez$^{c}$\\[6mm]}

{$^{a}$ \small Universidad San Sebastián, Facultad de Ingeniería,\\ [.01 cm]}
{\small\it Bellavista 7, Recoleta, Santiago, Chile.\\ [.3 cm]}
{$^{b}$ \small University of Wroclaw, Faculty of Physics and Astronomy,\\ [.01 cm]}
{\small\it Maksa Borna 9, 50-204 Wroclaw,
Poland\\ [.3 cm]}
{$^{c}$ \small Universidad Argentina de la Empresa (UADE), Facultad de Ingeniería y Ciencias Exactas, Departamento de Ciencias Básicas,\\ [.01 cm]}
{\small\it Lima 717, Buenos Aires, Argentina\\ [.5 cm]}

{\small \verb"ext.romina.ballester@uss.cl, eric.lescano@uwr.edu.pl, jesurodriguez@uade.edu.ar"}\\[1cm]

\small{\bf Abstract} \\[0.5cm]\end{center}

A covariant implementation of diffeomorphisms in the presence of local symmetries is a nontrivial aspect of gravitational theories. In Double Field Theory, this is achieved through the so-called generalized Kosmann derivative. In this work, we show that the generalized Kosmann derivative admits a natural formulation entirely in terms of generalized fluxes through the inclusion of a compensating term that plays the role of a generalized momentum map, yielding a fully determined and covariant operator that provides a covariant realization of generalized diffeomorphisms. When parameterized in terms of the field content of heterotic supergravity, the resulting symmetry transformations give rise to momentum maps at the supergravity level, offering a duality-covariant interpretation of these objects. This framework provides a natural setting for the construction of conserved currents and Noether charges in doubled geometry with internal symmetries, with direct implications for black hole thermodynamics and its higher-derivative corrections in a duality-covariant setting.

   
\newpage

\setcounter{page}{1}
\pagestyle{plain}
\renewcommand{\thefootnote}{\arabic{footnote}}
\setcounter{footnote}{0}

\tableofcontents

\newpage

\section{Introduction}

The standard notion of symmetry in gravity is provided by infinitesimal isometries, generated by vector fields $\xi$ that preserve the metric. This condition is expressed by the vanishing of the Lie derivative of the metric,
\be
\label{Lie-metric}
L_{\xi}g_{\mu\nu} = \nabla_{\mu}\xi_{\nu} + \nabla_{\nu}\xi_{\mu} = 0 \, ,
\ee
which defines Killing vectors. However, there are theories with non-trivial local symmetries, such as formulations based on vielbeins, for which the Lie derivative fails to fully capture the symmetry structure. In these cases, simply restoring covariance is not sufficient, since an isometry should preserve the entire geometric structure, including the orthonormal frame, in order to act consistently on frame-dependent fields. 

This motivates the search for an operator that is covariant under both diffeomorphisms and local gauge transformations, and whose action on the vielbein vanishes along isometries. For Lorentz symmetry, this is achieved by supplementing the Lie derivative with compensating terms that cancel its non-covariant contributions and ensure the invariance of field configurations along Killing flows. The resulting operator is the Kosmann derivative \cite{Lichnerowicz:1963, Kosmann:1966, Kosmann:1972, Ortin:2002qb}, defined for a vector $V_{\mu}{}^{a}$ as
\be
\label{Kosmann-derivative}
K_{\xi}V_{\mu}{}^{a} = L_{\xi}V_{\mu}{}^{a} + \xi^{\nu}w_{\nu b}{}^{a}V_{\mu}{}^{b} - \nabla^{[b}\xi^{a]}V_{\mu b}\, ,
\ee
where $w_{\mu ab}$ denotes the spin connection. When applied to the vielbein, it reduces to
\be
\label{Kosmann-vielbein}
K_{\xi}e_{\mu}{}^{a} = \frac{1}{2}e^{\nu a}L_{\xi}g_{\mu\nu}\, , 
\ee
which vanishes for any Killing vector, so that isometries preserve both the metric and the orthonormal frame. This construction extends straightforwardly to general Lorentz tensors with arbitrary spacetime and tangent-space indices \cite{Ortin:2002qb}, providing the correct transformation properties for all fields in the theory.

These ideas were further applied in the context of black hole thermodynamics \cite{Jacobson:2015uqa, Prabhu:2015vua, Edelstein:2018ewc, Edelstein:2019wzg, Aneesh:2020fcr, Elgood:2020svt, Elgood:2020mdx, Ortin:2022uxa, Ballesteros:2023iqb, Ballesteros:2023muf, Ballesteros:2024prz, Ballesteros:2025wvs}, where covariance of the Lie derivative with respect to the symmetries of the theory plays a central role in the construction of Noether charges. In this context, in order to accommodate the field content of different supergravity theories, gauge-covariant Lie derivatives can be constructed using the so-called momentum maps \cite{Elgood:2020mdx, Elgood:2020svt}. Given a spacetime vector $\xi$, the associated momentum map $P_{\xi}$ plays a central role in the compensating gauge transformation required so that the combined action of diffeomorphisms and local symmetries acts consistently on the fields. From this perspective, covariant symmetry operators, such as the Kosmann derivative, arise naturally from the underlying geometric structure, rather than as ad hoc modifications of the standard Lie derivative. 

This viewpoint allows one to reinterpret the Kosmann derivative as the gauge-covariant realization of an isometry acting on fields with Lorentz indices. In the case of the vielbein, Lorentz covariance and invariance along Killing flows are simultaneously ensured by supplementing the diffeomorphism with a compensating local Lorentz transformation parameterized by
\be
\Lambda_{\xi}^{ab} = \xi^{\mu}w_{\mu}{}^{ab} - P_{\xi}^{ab}\, ,
\ee
which leads to the gauge-covariant transformation
\be
\label{vielbein-gauge-covariant}
\delta_{\xi}e_{\mu}{}^{a} = L_{\xi}e_{\mu}{}^{a} + \Lambda_{\xi b}{}^{a}e_{\mu}{}^{b} = \nabla_{\mu}\xi^{a} - \nabla^{[b}\xi^{a]}e_{\mu b}\, ,
\ee
where $P_{\xi}^{ab}=\nabla^{[a}\xi^{b]}$ is the momentum map associated with $\xi$. As a result, $\delta_{\xi}$ coincides precisely with the Kosmann derivative $K_{\xi}$ in \eqref{Kosmann-derivative}.

Momentum maps can be defined for more general fields and gauge symmetries \cite{Elgood:2020svt, Elgood:2020mdx}, including fields such as the Kalb–Ramond two-form of the NS–NS sector of supergravity theories, whose gauge structure cannot be straightforwardly accommodated within more restrictive frameworks such as \cite{Prabhu:2015vua}. This generality makes the momentum-map approach particularly well suited for supergravity and string-inspired scenarios, where spacetime and gauge symmetries are deeply intertwined. A natural question is how this approach behaves in the presence of string dualities, and in particular under T-duality. Understanding how covariant symmetry operators act in a duality-covariant setting therefore requires a framework in which T-duality is realized manifestly, and where diffeomorphisms and gauge transformations are treated on equal footing from the outset.

Double Field Theory (DFT) \cite{Siegel:1993xq, Siegel:1993th, Hull:2009mi, Hohm:2010pp} provides such a framework, unifying spacetime diffeomorphisms and gauge transformations into generalized diffeomorphisms acting on an extended space. Additional local symmetries, such as Lorentz symmetry, are realized through double Lorentz transformations \cite{Hohm:2010xe, Hohm:2011ex}. This structure naturally calls for a covariant notion of symmetry transformations acting on fields carrying (double-)local indices, leading to the introduction of a generalized notion of the Kosmann derivative, which will be discussed next. 

\subsection{State of the art and main results}

The generalized Kosmann derivative was originally constructed in \cite{Angus:2018mep} within the semi-covariant formulation of DFT, and more recently reviewed in \cite{Park:2025ugx}. In this approach, the doubled geometry is fully specified at the price of introducing a connection that transforms only semi-covariantly. As in ordinary Riemannian geometry, the generalized Kosmann derivative characterizes isometries through the vanishing of its action on the generalized frame, providing a consistent operator which preserves the double Lorentz symmetry.

Since many DFT constructions require the flux formulation \cite{Hohm:2010xe}, such as the higher-derivative corrections discussed in Section \ref{hdc}, it is desirable to extend the prescription for the generalized Kosmann derivative introduced in \cite{Angus:2018mep} to the flux formalism. While at first sight this could affect the determined nature of the generalized Kosmann derivative, we show in this paper that that this is not the case: it is possible to construct a compensating term, which plays the role of a generalized momentum map and allows one to fully determined the generalized Kosmann derivative in the flux formulation of DFT. This extension is particularly relevant for applications to conserved charges \cite{Blair:2015eba, Naseer:2015fba, Park:2015bza} and black hole thermodynamics \cite{Arvanitakis:2016zes}, where symmetry generators and their associated momentum maps play a central role. Our construction provides a manifestly duality-covariant framework to analyze such questions in the presence of both spacetime and internal symmetries.

The main results of this work can be summarized as follows:

\begin{itemize}

\item We show that the generalized Kosmann derivative can be expressed entirely in terms of generalized fluxes by introducing a compensating term that plays the role of a generalized momentum map (sections \ref{ss-3.1} and \ref{ss-3.2}). This provides a fully covariant construction within the flux formulation of DFT, extending the results of \cite{Angus:2018mep}.

\item We demonstrate that symmetry transformations generated by the generalized Kosmann derivative naturally give rise to momentum maps in heterotic supergravity, as presented in section \ref{subsection-MM}. In particular, upon identifying the generalized diffeomorphism parameters with the corresponding momentum maps, the resulting transformations precisely reproduce those obtained in \cite{Elgood:2020mdx, Elgood:2020svt}, thereby providing a manifestly duality covariant interpretation of those objects.

\item We analyze the geometrical properties of the object ${\cal P}_{\xi}^{AB}$ and its relation to the generalized Riemann tensor ${\cal R}_{MNPQ}$. While in Riemannian geometry the momentum map satisfies
\be
\nabla_{\mu}P_{\xi}^{ab} = - \xi^{\nu}R_{\nu\mu}{}^{ab}\, ,
\ee
we show that its generalization to doubled geometry involves additional terms, given explicitly in \eqref{MomentumrelationDFT}, which are required in order to ensure manifest duality invariance.

\end{itemize}

\subsection{Organization of the paper}

This paper is organized as follows. In Section 2 we present a brief review of DFT, fixing our notation and conventions. In Section 3 we discuss in detail the construction of the generalized Kosmann derivative within the flux formulation of DFT. There, we parameterize the resulting expressions in terms of the field content of heterotic supergravity and compare them with the momentum map formulation introduced in \cite{Elgood:2020mdx, Elgood:2020svt}. The section closes with a discussion of the geometrical properties of momentum maps in both DFT and supergravity. Section 4 is devoted to explore the implications of our results for the construction of Noether charges and their relevance for black hole thermodynamics in a manifestly duality-covariant framework, as well as for the inclusion of higher-derivative corrections. Finally, Section 5 contains our conclusions and outlook.

\section{Double Field Theory: a short pedagogical review}

In this section, we briefly review the aspects of DFT that will be used along the paper. More comprehensive reviews can be found in \cite{Aldazabal:2013sca, Berman:2013eva, Hohm:2013bwa, Park:2025ugx}.

\subsection{Generalized metric formulation}

DFT provides a framework that makes the T-duality symmetry of string theory manifest at the level of a field theory. It is formulated on a doubled space with coordinates $X^{M}=\left(x^{\mu},\tilde{x}_{\mu}\right)$, which transform in the fundamental representation of the $O(D,D)$ symmetry group. The term \textit{doubled} refers to the fact that the spacetime dimension is extended from $D$ to $2D$, introducing dual coordinates that account for winding excitations of strings. The indices $M, N, \dots = 0, 1, \dots, 2D-1$ label the doubled coordinates, while $\mu, \nu, \dots = 0, 1, \dots, D-1$ correspond to the physical subspace. This doubling enables a manifestly symmetric description of both momentum and winding modes.

The fields of DFT are organized into multiplets of the duality group, making $O(D,D)$ a manifest global symmetry. The group preserves an invariant metric $\eta_{MN}$, defined as
\be
\label{O(D,D)-invariant}
\eta_{MN} = \left(\begin{matrix} 0 & \delta^{\mu}{}_{\nu}\\ 
\delta_{\mu}{}^{\nu} & 0 \end{matrix}\right)\, , 
\ee
which, together with its inverse $\eta^{MN}$, is used to raise and lower indices in the doubled space. In particular, the fundamental fields of the framework are the generalized metric $H_{MN}$, a symmetric $O(D,D)$ tensor, and the generalized dilaton $d$. Together they encode the massless bosonic sector of string theory through the parameterization
\be
\label{generalizedFields}
H_{MN} = \begin{pmatrix} g^{\mu \nu} & - g^{\mu \rho} b_{\rho \nu} \\ b_{\mu \rho} g^{\rho \nu} & g_{\mu \nu} - b_{\mu \rho} g^{\rho \sigma} b_{\sigma \nu}\end{pmatrix}\, , \qquad e^{-2d} = \sqrt{-g}e^{-2\phi}\, ,
\ee
where $g_{\mu\nu}$ is the $D$-dimensional metric, $b_{\mu\nu}$ the Kalb–Ramond two-form, $g$ the determinant of the metric $g_{\mu\nu}$, and $\phi$ the dilaton field. The combination $e^{-2d}$ is a scalar of weight one which will be used as an integration measure.

It is possible to introduce a natural decomposition of the doubled space into two orthogonal subspaces by combining the $O(D,D)$-invariant metric and the generalized metric. This is achieved through the projectors
\be
\label{dft-projectors}
P_{MN} = \frac{1}{2}\left(\eta_{MN} - H_{MN}\right)\, , \qquad \ov{P}_{MN} = \frac{1}{2}\left(\eta_{MN} + {H}_{MN}\right)\ ,
\ee
which allow tensors to be consistently separated into components as
\be
V_{M} \ = \ P_{M}{}^{N}V_{N} + \ov{P}_{M}{}^{N}V_{N} \ = \ V_{\un{M}} + V_{\ov{M}}\, .
\ee

Beyond the coordinate doubling, DFT also extends local symmetries to reflect the dualities inherited from string theory. In particular, generalized diffeomorphisms act on tensor densities through the generalized Lie derivative, defined as
\be
\label{generalized-Lie-derivative}
{\cal{L}}_{\xi}V^{M} = \xi^{N}\partial_{N}V^{M} + \left(\partial^{M}\xi_{N} - \partial_{N}\xi^{M}\right)V^{N} + w(V)\partial_{N}\xi^{N}V^{M}\, ,
\ee
where $\xi^{M} = (\xi^{\mu}, \lambda_{\mu})$ is the parameter of generalized diffeomorphisms. The components $\xi^{\mu}$ correspond to conventional diffeomorphisms in $D$-dimensional spacetime, while $\lambda_{\mu}$ generates abelian gauge transformations of the Kalb–Ramond field. Closure of the algebra of generalized diffeomorphisms requires the strong constraint,
\be
\label{dft-strong-constraint}
\partial_{M}\partial^{M}(\cdots) \ = \ \partial_{M}(\cdots)\partial^{M}(\cdots) \ = \ 0 \, ,
\ee
where $(\cdots)$ denotes generalized fields or their products. This condition restricts the coordinate dependence of all fields and gauge parameters, effectively reducing the theory to a $D$-dimensional physical subspace while preserving $O(D,D)$ covariance. 

Generalized diffeomorphism covariance can be formulated in terms of a covariant derivative
\be 
\label{covariant-derivative}
\nabla_{M}V_{N} = \partial_{M}V_{N} - \Gamma_{MN}{}^{P}V_{P}\, ,
\ee
with $\Gamma_{MN}{}^{P}$ the generalized affine connection. As is well known, this connection is not completely fixed by symmetry requirements alone.

The dynamics of DFT is governed by an action invariant under both generalized diffeomorphisms and global $O(D,D)$ transformations. It is given by
\be 
\label{DFT-action-metric} 
S = \int d^{2D}X \ e^{-2d}{\cal R}(H,d)\, .
\ee
where the generalized Ricci scalar ${\cal R}(H,d)$ is defined as
\bea 
\label{generalized-Ricci-H} 
{\cal R}(H,d) & = & 4H^{MN}\partial_{M}\partial_{N}d - \partial_{M}\partial_{N}H^{MN} + 4\partial_{M}H^{MN}\partial_{N}d - 4H^{MN}\partial_{M}d\partial_{N}d\, \nn \\ 
& & - \frac{1}{2}H^{MN}\partial_{M}H^{KL}\partial_{K}H_{NL} + \frac{1}{8}H^{MN}\partial_{M}H^{KL}\partial_{N}H_{KL}\, .
\eea 
By parameterizing the fundamental fields in terms of supergravity multiplets and solving the strong constraint, the action \eqref{DFT-action-metric} reduces to the NS–NS supergravity effective action.

\subsection{Flux formulation}

The introduction of a vielbein in gravity theories is central to describe local internal symmetries, such as Lorentz transformations, and to explore the local structure of the spacetime, providing a map between the tangent space and the manifold. In DFT, this role is played by the generalized frame.

In this formulation, the fundamental field is not the generalized metric but the generalized frame $E_{M}{}^{A}$, which parameterizes the coset $O(D,D)/O(D-1,1)_{L}\times O(1,D-1)_{R}$, where the local group $O(D-1,1)_{L}\times O(1,D-1)_{R}$ represents the duplication of the Lorentz symmetry. The generalized frame satisfies the conditions
\be
\label{dft-frame-constraints}
\eta_{MN} = E_{M}{}^{A}\eta_{AB}E_{N}{}^{B}\, , \qquad H_{MN} = E_{M}{}^{A}H_{AB}E_{N}{}^{B}\, ,
\ee
where the flat indices $A$ decompose as $A=\left(\un{a},\ov{a}\right)$, with $\un{a},\ov{a}=0,1,\dots,D-1$. The flat metrics $\eta_{AB}$ and $H_{AB}$ 
\be
\label{dft-h-metrics}
\eta_{AB} = \left(\begin{matrix} -g_{\un{ab}} & 0\\
0 & g_{\ov{ab}}\end{matrix}\right)\, , \quad H_{AB} = \left(\begin{matrix} g_{\un{ab}} & 0\\
0 & g_{\ov{ab}}\end{matrix}\right)\, , \quad g_{\un{ab}} = g_{\ov{ab}} = \text{diag}(-1, \underbrace{+1, \dots, +1}_{D-1 \text{ times}})\, ,
\ee
are constants, invertibles, and invariants under the double Lorentz group. The metric $\eta_{AB}$ is used to raise and lower flat indices. 

The generalized frame can be parameterized in terms of the $D$-dimensional vielbeins and the Kalb–Ramond field as
\be
\label{dft-generalized_frame}
E_{M}{}^{A} = \frac{1}{\sqrt{2}}\left(\begin{matrix} e^{\mu}{}_{\un{a}} & e^\mu{}_{\ov{a}} \\
- e_{\mu\un{a}} + b_{\mu\nu}e^{\nu}{}_{\un{a}} \  &  \ e_{\mu\ov{a}} + b_{\mu\nu}e^{\nu}{}_{\ov{a}}\end{matrix}\right)\, ,
\ee
The $D$-dimensional vielbeins $e_{\mu}{}^{\un{a}}$ and $e_{\mu}{}^{\ov{a}}$ satisfy
\be
\label{dft-nd_vielbein}
e_{\mu}{}^{\un{a}}g_{\un{ab}}e_{\nu}{}^{\un{b}} = e_{\mu}{}^{\ov{a}}g_{\ov{ab}}e_{\nu}{}^{\ov{b}} = g_{\mu\nu}\, ,
\ee
and, after an appropriate gauge fixing, they can be identified as $e_{\mu}{}^{\un{a}}\delta_{\un{a}}^{a}=e_{\mu}{}^{\ov{a}}\delta_{\ov{a}}^{a}=e_{\mu}{}^{a}$, where $e_{\mu}{}^{a}$ is the supergravity vielbein, and the indices $a,b,\dots=0,\dots,D-1$ belong to $O(1,D-1)$. This identification reduces the double Lorentz group to the conventional Lorentz group.

Just as in the formulation based on the generalized metric, it is possible to decompose the doubled tangent space by projecting flat indices with combinations of the invariant flat metrics. This leads to the definition of the flat projectors
\be
\label{flat_projectors}
P_{AB} = \frac{1}{2}\left(\eta_{AB} - H_{AB}\right)\, , \qquad \ov{P}_{AB} = \frac{1}{2}\left(\eta_{AB} + {H}_{AB}\right)\ ,
\ee
which allow one to decompose a double Lorentz vector into its projected components
\be
V_{A} = P_{A}{}^{B}V_{B} + \ov{P}_{A}{}^{B}V_{B}\, .
\ee
These components represent the projections of the vector onto the subspaces defined by the projectors, corresponding to the two copies of the double Lorentz group. The projectors and their inverses are used to raise and lower projected indices.

In addition to generalized diffeomorphisms, Lorentz transformations must also be extended to double space. An infinitesimal double Lorentz transformation acting on a generic vector $V_{A}$ takes the form
\be
\label{double-Lorentz}
\delta_{\Lambda}V_{A} = V_{B}\Lambda^{B}{}_{A}\, .
\ee
Requiring the flat metrics $\eta_{AB}$ and $H_{AB}$ to remain invariant under this transformation imposes conditions on the double Lorentz parameter $\Lambda_{AB}$. Invariance of $\eta_{AB}$ yields the antisymmetry condition $\Lambda_{AB}=-\Lambda_{BA}$, while invariance of $H_{AB}$ leads to $\Lambda_{\un{a}\ov{b}}=\Lambda_{\ov{a}\un{b}}=0$, ensuring that the transformations act independently on the two Lorentz sectors.

As usual, the presence of local symmetries requires introducing covariant derivatives. In the presence of double Lorentz symmetry, the covariant derivative \eqref{covariant-derivative} takes the form
\be
\label{dft-covariant}
\nabla_{M}V_{N A} = \partial_{M}V_{N A} - \Gamma_{MN}{}^{P}V_{P A} - \omega_{MA}{}^{B}V_{N B}\, ,
\ee
where $\omega_{MA}{}^{B}$ is the generalized spin connection, constrained by
\be
\omega_{MAB} = -\omega_{MBA}\, , \qquad \omega_{M\un{a}\ov{b}} = \omega_{M\ov{a}\un{b}} = 0\, .
\ee
Unlike in standard Riemannian geometry, not all components of the generalized spin connection are determined in DFT. However, imposing compatibility of the covariant derivative with the generalized frame, $\nabla_{M}E_{N}{}^{A}=0$, together with vanishing torsion, allows one to define the generalized fluxes in terms of fully determined components of the spin connection, namely
\be
\label{dft-generalized-fluxes}
F_{ABC} = 3\omega_{[ABC]} = 3\partial_{[A}E^{M}{}_{B}E_{|M|C]}\, , \qquad F_{A} = \omega_{BA}{}^{B} = \sqrt{2}e^{2d}\partial_{M}\left(E^{M}{}_{A}e^{-2d}\right)\, ,
\ee
where $\partial_{A}=\sqrt{2}E^{M}{}_{A}\partial_{M}$ is the flat derivative. These fluxes play a key role in encoding the dynamics of the theory. 

In terms of the generalized frame, the action \eqref{DFT-action-metric} can be rewritten as
\be
\label{dft-action}
S = \int d^{2D}X \ e^{-2d}{\cal{R}}(E,d)\, ,
\ee
where the generalized Ricci scalar $\mathcal{R}$ is expressed in terms of the fluxes as
\be
\label{dft-generalized_ricci}
{\cal{R}} = 2\partial_{\un{a}}F^{\un{a}} + F_{\un{a}}F^{\un{a}} - \frac{1}{6}F_{\un{abc}}F^{\un{abc}} - \frac{1}{2}F_{\ov{a}\un{bc}}F^{\ov{a}\un{bc}}\, .
\ee
Upon parameterizing the generalized frame and dilaton, this action reproduces the standard NS–NS sector of supergravity, in complete analogy with the formulation based on the generalized metric.

\subsection{Heterotic extension}

DFT can be naturally extended to incorporate the degrees of freedom associated with (super) Yang-Mills fields \cite{Hohm:2011ex}. In this case, the global duality symmetry group is enlarged to $O(D,D+n_g)$, where $n_g$ denotes the dimension of the gauge group.\footnote{For $n_g=496$, this construction corresponds to heterotic DFT, as it matches the dimension of the gauge groups $SO(32)$ and $E_8\times E_8$.} Accordingly, the generalized coordinates are extended from $2D$ to $2D+n_g$,
\be
X^{M} = \left(x^{\mu},\tilde{x}_{\mu}, x^{i}\right)\, , \qquad i={1,\dots,n_{g}}\, .
\ee

The tangent-space symmetry group is also enlarged to
$O(D,1)_{L}\times O(1,D+n_{g})_{R}$, implying the decomposition of flat indices
\be
A = (\un{a},\ov{A}) = (\un{a},\ov{a}, \ov{i})\, , \qquad \ov{i} = 1,\dots,n_{g}\, .
\ee
The invariant metrics then take the form
\be
\eta_{MN} = \left(\begin{matrix} 0 & \delta^{\mu}{}_{\nu} & 0\\
\delta_{\mu}{}^{\nu} & 0 & 0\\
0 & 0 & \kappa_{ij}\end{matrix}\right)\, , \quad \eta_{AB} = \left(\begin{matrix} - g_{\un{ab}} & 0 & 0\\
0 & g_{\ov{ab}} & 0\\
0 & 0 & \kappa_{\ov{ij}}\end{matrix}\right)\, , \quad H_{AB} = \left(\begin{matrix} g_{\un{ab}} & 0 & 0\\
0 & g_{\ov{ab}} & 0\\
0 & 0 & \kappa_{\ov{ij}}\end{matrix}\right)\, ,
\ee
where $\kappa_{ij}=e_{i}{}^{\ov{i}}\kappa_{\ov{ij}}e_{j}{}^{\ov{j}}$ is the Cartan-Killing metric of the gauge group, and $e_{i}{}^{\ov{i}}$ plays the role of a vielbein on the gauge algebra.\footnote{Can also be interpreted as the scalar fields arising from compactifications.}

Non-abelian gauge interactions enter through the gaugings $f_{MNP}$ \cite{Grana:2012rr}, parameterized as
\bea
\label{fidentification}
f_{MNP} = \left\{\begin{matrix}f_{ijk} & {\rm for} \ \ {MNP}=i,j,k \\
0 & {\rm other} \ {\rm combinations}\end{matrix}\right.\, , 
\eea
where $f_{ijk}$ are the structure constants of the non-abelian gauge group. These gauging contributions deform both the generalized Lie derivative,
\be
\label{gauged-Lie-derivative}
{\cal{L}}_{\xi}V^{M} = \xi^{N}\partial_{N}V^{M} + \left(\partial^{M}\xi_{N} - \partial_{N}\xi^{M}\right)V^{N} + w(V)\partial_{N}\xi^{N}V^{M} + f^{M}{}_{NP}\xi^{N}V^{P}\, ,
\ee
where the generalized parameter $\xi_{M}=\left(\xi^{\mu},\lambda_{\mu},\chi^{i}\right)$ is extended to include the gauge parameter $\chi^{i}$, and the generalized fluxes
\be
\label{gauged-fluxes}
F_{ABC} = 3\partial_{[A}E^{M}{}_{B}E_{|M|C]} + \sqrt{2}E^{M}{}_{A}E^{N}{}_{B}E^{P}{}_{C}f_{MNP}\, .
\ee
The strong constraint \eqref{dft-strong-constraint} is also modified in the presence of gaugings,
\be
\label{strong-constraint-gauge}
\partial_{M}\partial^{M}(\cdots) = 0\, , \qquad \partial_{M}(\cdots)\partial^{M}(\cdots) = 0\, , \qquad f_{MN}{}^{P}\partial_{P}(\cdots) = 0\, ,
\ee
supplemented by the quadratic constraints
\be
\label{strong-constraint-gauge-quadratic}
f_{MNP} = f_{[MNP]}\, , \qquad f_{[MN}{}^{R}f_{P]R}{}^{Q} = 0\, .
\ee

The projection of the double Lorentz parameter must be extended to $\Lambda_{\ov{AB}}=(\Lambda_{\ov{ab}},\Lambda_{\ov{ai}},\Lambda_{\ov{ij}})$, with the gauge-index components fixed consistently with the chosen parameterization in order to avoid introducing spurious degrees of freedom. In that sense, the generalized frame $E^{M}{}_{A}$ can be parameterized as
\be
\label{frame-gauge-param}
E_{M}{}^{A} = \frac{1}{\sqrt{2}}\begin{pmatrix} e^{\mu}{}_{\un{a}} & e^{\mu}{}_{\ov{a}} & 0\\
- e_{\mu\un{a}} - C_{\nu\mu}e^{\nu}{}_{\un{a}} & e_{\mu\ov{a}} - C_{\nu\mu}e^{\nu}{}_{\ov{a}} & \sqrt{2}A_{\mu i}e^{i}{}_{\ov{i}}\\
- A_{\mu}{}^{i}e^{\mu}{}_{\un{a}} & - A_{\mu}{}^{i}e^{\mu}{}_{\ov{a}} & \sqrt{2}e^{i}{}_{\ov{i}} \end{pmatrix}\, ,
\ee
where $A_{\mu}{}^{i}$ are gauge vectors and $C_{\mu\nu}=b_{\mu\nu}+\tfrac{1}{2}A_{\mu}{}^{i}A_{\nu i}$. 

The corresponding action is obtained by extending \eqref{dft-action} to the enlarged duality group, which amounts to including the gauge-sector contributions in the generalized Ricci scalar \eqref{dft-generalized_ricci}. Upon parametrization in terms of supergravity fields, this action reproduces the NS-NS sector coupled to Yang-Mills fields. In particular, for gauge groups $SO(32)$ or $E_{8}\times E_{8}$, this reproduces the low-energy effective action of heterotic string theory.

\subsection{Isometries in Double Field Theory}

Isometries in DFT are defined as generalized diffeomorphisms that leave invariant both the generalized metric and the generalized dilaton,
\be
\label{isometries}
{\cal L}_{\xi}H_{MN} = 0\, , \qquad {\cal L}_{\xi}d = 0\, ,
\ee
extending the notion of Killing symmetries in ordinary geometry to double space.

Let us first consider the generalized metric $H_{MN}$. Its transformation under generalized diffeomorphisms is given by
\be
{\cal L}_{\xi}H_{MN} = \xi^{Q}\partial_{Q}H_{MN} + \left(\partial_{M}\xi^{Q} - \partial^{Q}\xi_{M}\right)H_{QN} + \left(\partial_{N}\xi^{Q} - \partial^{Q}\xi_{N}\right)H_{MQ}\, ,
\ee
which can be conveniently rewritten in terms of covariant derivatives as
\be
{\cal L}_{\xi}H_{MN} = \left(\nabla_{M}\xi^{Q} - \nabla^{Q}\xi_{M}\right)H_{QN} + \left(\nabla_{N}\xi^{Q} - \nabla^{Q}\xi_{N}\right)H_{MQ}\, .
\ee
Using the projectors \eqref{dft-projectors}, this result can be cast into the compact form
\be
{\cal L}_{\xi}H_{MN} = 8P_{(M}{}^{Q}\ov{P}_{N)}{}^{R}\nabla_{[Q}\xi_{R]}\, .
\ee

For the generalized dilaton, the generalized Lie derivative reads
\be
{\cal L}_{\xi}d = \xi^{M}\partial_{M}d - \frac{1}{2}\partial_{M}\xi^{M} = - \frac{1}{2}\nabla_{M}\xi^{M}\, .
\ee
where in the last step we have used the identity $\Gamma_{NM}{}^{N}=-2\partial_{M}d$, which is fixed by consistency of the covariant derivative with the invariant measure $e^{-2d}$. Therefore, a generalized vector $\xi^{M}$ generates an isometry of double space if and only if it satisfies the generalized Killing equations \cite{Park:2015bza}
\be
\label{DFT-killing}
P_{(M}{}^{Q}\ov{P}_{N)}{}^{R}\nabla_{[Q}\xi_{R]} = 0\, , \qquad \nabla_{M}\xi^{M} = 0\, .
\ee

Under generalized diffeomorphisms, the generalized frame $E_{M}{}^{A}$ transforms as
\be
\label{lie-frame-1}
{\cal L}_{\xi}E_{M}{}^{A} = \xi^{N}\partial_{N}E_{M}{}^{A} + \left(\partial_{M}\xi^{N} - \partial^{N}\xi_{M}\right)E_{N}{}^{A}\, ,
\ee
which, when written in terms of covariant derivatives, becomes
\be
\label{lie-frame-2}
{\cal L}_{\xi}E_{M}{}^{A} = \left(\nabla_{M}\xi^{N} - \nabla^{N}\xi_{M}\right)E_{N}{}^{A} - \xi^{N}\omega_{NB}{}^{A}E_{M}{}^{B}\, .
\ee
This expression is neither covariant under double Lorentz transformations nor preserved by generalized isometries. This signals that the generalized Lie derivative must be further extended in order to incorporate local Lorentz covariance and to define consistent Killing equations for the generalized frame.

\section{Momentum maps from Double Field Theory}
\label{section-3}

In this section we establish the relation between the generalized Kosmann derivative in DFT and the notion of momentum maps arising in heterotic supergravity. We first construct the generalized Kosmann derivative within the flux formulation of DFT, identifying the compensating term required to ensure double Lorentz covariance and compatibility with generalized isometries. We then parametrize the resulting transformations in terms of the bosonic fields of heterotic supergravity, where the compensating term acquires a natural interpretation as a momentum map. The section concludes with a discussion of the geometrical properties of this object in the doubled framework.

\subsection{The generalized Kosmann derivative in flux formulation}
\label{ss-3.1}
The analysis of isometries in DFT shows that, although isometries are well defined for the generalized metric and dilaton, the generalized Lie derivative does not act covariantly on the generalized frame. This mirrors the situation in conventional gravity and motivates the construction of a Lorentz-covariant extension of the generalized Lie derivative, namely the generalized Kosmann derivative.

Consider a generalized vector $V_{M}{}^{A}$ (we take $w(V)=0$ for simplicity). Under an infinitesimal double Lorentz transformation \eqref{double-Lorentz}, its generalized Lie derivative transforms as
\be
\delta_{\Lambda}\left({\cal{L}}_{\xi}V_{M}{}^{A}\right) = {\cal L}_{\xi}V_{M}{}^{B}\Lambda_{B}{}^{A} + \xi^{N}\partial_{N}\Lambda_{B}{}^{A}V_{M}{}^{B}\, .
\ee
The second term on the right-hand side signals the lack of covariance of the generalized Lie derivative under double Lorentz. Then, the standard notion of Lie transport along vector fields is therefore not compatible with the local symmetries of DFT. 

To restore Lorentz covariance, the generalized Lie derivative must be supplemented by an additional term whose Lorentz variation cancels the non-covariant contribution above. A natural candidate is
\be
\xi^{N}\omega_{NB}{}^{A}V_{M}{}^{B}\, , \nn
\ee
which transforms precisely in the required way. This leads to a Lorentz-covariant generalized Lie derivative
\be
\label{Lie-Lorentz-der}
{\cal L}_{\xi}^{(\nabla)}V_{M}{}^{A} = {\cal L}_{\xi}V_{M}{}^{A} + \xi^{N}\omega_{NB}{}^{A}V_{M}{}^{B} = \xi^{N}\nabla_{N}V_{M}{}^{A} + \left(\nabla_{M}\xi^{N} - \nabla^{N}\xi_{M}\right)V_{N}{}^{A}\, .
\ee
Although this operator is, by construction, covariant under double Lorentz transformations, it still suffers from two limitations. First, its action on the generalized frame does not vanish for generalized Killing vectors, showing that Lorentz covariance alone is insufficient to characterize isometries at the frame level. Second, it depends on undetermined components of the generalized spin connection, obstructing contact with the underlying $D$-dimensional physics.

We therefore seek an operator
\be
\label{Kosmann-schem}
{\cal K}_{\xi} = {\cal L}_{\xi}^{(\nabla)} + {\cal P}_{\xi}\, ,
\ee
that preserves both double Lorentz covariance and generalized isometries. Lorentz covariance requires the compensating term ${\cal P}_{\xi}$ to be itself a double Lorentz transformation constructed from the generalized diffeomorphism parameter, acting on double Lorentz indices. Its action on a Lorentz tensor is analogous to that generated by $\Lambda_{AB}$, namely
\be
{\cal P}_{\xi}V^{A} = V^{B}{\cal P}_{\xi B}{}^{A}\, ,
\ee
with ${\cal P}_{\xi}$ satisfying 
\be
\label{conditions-P}
{\cal P}_{\xi AB} = - {\cal P}_{\xi BA}\, , \qquad {\cal P}_{\xi \ov{a}\un{b}} = {\cal P}_{\xi \un{a}\ov{b}} = 0\, .
\ee
Compatibility with generalized isometries further imposes
\be
{\cal K}_{\xi}E_{M}{}^{\un{a}} = \alpha E^{N \un{a}}{\cal L}_{\xi}H_{MN}\, , \qquad {\cal K}_{\xi}E_{M}{}^{\ov{a}} = \beta E^{N \ov{a}}{\cal L}_{\xi}H_{MN}\, ,
\ee
and our task is then to determine the explicit form of ${\cal P}_{\xi}$ consistent with these conditions. 

A straightforward computation yields
\bea
E_{M B}{\cal P}^{B\un{a}} & = & \frac{1}{\sqrt{2}}\left(2\alpha\nabla^{\un{a}}\xi^{\ov{b}} - 2\alpha\nabla^{\ov{b}}\xi^{\un{a}} - \nabla^{\ov{b}}\xi^{\un{a}} + \nabla^{\un{a}}\xi^{\ov{b}}\right)E_{M\ov{b}} - \frac{1}{\sqrt{2}}\left(\nabla^{\un{b}}\xi^{\un{a}} - \nabla^{\un{a}}\xi^{\un{b}}\right)E_{M\un{b}}\, , \nn \\
E_{M B}{\cal P}^{B\ov{a}} & = & \frac{1}{\sqrt{2}}\left(2\beta\nabla^{\un{b}}\xi^{\ov{a}} - 2\beta\nabla^{\ov{a}}\xi^{\un{b}} - \nabla^{\un{b}}\xi^{\ov{a}} + \nabla^{\ov{a}}\xi^{\un{b}}\right)E_{M\un{b}} - \frac{1}{\sqrt{2}}\left(\nabla^{\ov{b}}\xi^{\ov{a}} - \nabla^{\ov{a}}\xi^{\ov{b}}\right)E_{M\ov{b}}\, \nn .
\eea
Since mixed components must vanish, the coefficients $\alpha$ and $\beta$ are fixed to be\footnote{Which agrees with the standard Kosmann derivative \eqref{Kosmann-vielbein} acting on the vielbein upon reduction to $D$-dimensional supergravity.}
\be
{\cal K}_{\xi}E_{M}{}^{\un{a}} = - \frac{1}{2}E^{N \un{a}}{\cal L}_{\xi}H_{MN}\, , \qquad {\cal K}_{\xi}E_{M}{}^{\ov{a}} = \frac{1}{2}E^{N \ov{a}}{\cal L}_{\xi}H_{MN}\, , 
\ee
which in turn determines the compensating parameter
\be
\label{generalized-momentum-map}
{\cal P}_{\xi AB} = - \sqrt{2}\nabla_{[A}\xi_{B]}\, ,
\ee
satisfying conditions \eqref{conditions-P}.

With these ingredients in place, we can now present the generalized Kosmann derivative acting on a Lorentz vector
\be
\label{genKosmann}
{\cal K}_{\xi}V_{M}{}^{A} = \xi^{N}\nabla_{N}V_{M}{}^{A} + \left(\nabla_{M}\xi^{N} - \nabla^{N}\xi_{M}\right)V_{N}{}^{A} - \sqrt{2}\nabla^{[\un{b}}\xi^{\un{a}]}V_{M\un{b}} - \sqrt{2}\nabla^{[\ov{b}}\xi^{\ov{a}]}V_{M\ov{b}}\, ,
\ee
in terms of the diffeomorphism parameter $\xi^{M}$. At this stage, it is not manifest that the operator is fully determined. To make this explicit, we rewrite
\be
{\cal K}_{\xi}V_{M}{}^{A} = {\cal L}_{\xi}V_{M}{}^{A} + \xi^{N}\omega_{N\un{b}}{}^{\un{a}}V_{M}{}^{\un{b}} + \xi^{N}\omega_{N\ov{b}}{}^{\ov{a}}V_{M}{}^{\ov{b}} - \sqrt{2}\nabla^{[\un{b}}\xi^{\un{a}]}V_{M\un{b}} - \sqrt{2}\nabla^{[\ov{b}}\xi^{\ov{a}]}V_{M\ov{b}}\, .
\ee
Flattening the first index of the generalized spin connection according to $\omega_{ABC}=\sqrt{2}E^{M}{}_{A} \omega_{MBC}$, and expanding the covariant derivatives in the last two terms, we find
\be
\label{kosmann-determined}
{\cal K}_{\xi}V_{M}{}^{A} = {\cal L}_{\xi}V_{M}{}^{A} - \sqrt{2}\partial^{[\un{b}}\xi^{\un{a}]}V_{M\un{b}} - \sqrt{2}\partial^{[\ov{b}}\xi^{\ov{a}]}V_{M\ov{b}} - \frac{1}{\sqrt{2}}\xi^{C}F_{C}{}^{\un{ab}}V_{M\un{b}} - \frac{1}{\sqrt{2}}\xi^{C}F_{C}{}^{\ov{ab}}V_{M\ov{b}}\, .
\ee
Since the generalized fluxes are fully determined, this expression shows that the generalized Kosmann derivative is completely fixed.

The closure of the generalized Kosmann operator is given by the following expressions, 
\be
\left[{\cal K}_{\xi_2}, {\cal K}_{\xi_1}\right] V_{M}{}^{\un a} = {\cal K}_{\xi_{12}} V_{M}{}^{\un a} + V_{M \un b}\Lambda_{12}^{\un{ba}}\, , \qquad \left[{\cal K}_{\xi_2}, {\cal K}_{\xi_1}\right]V_{M}{}^{\overline a} = {\cal K}_{\xi_{12}} V_{M}{}^{\overline a} + V_{M \overline b}\Lambda_{12}^{\overline{ba}}\, ,
\ee
where $\xi_{12}^{M}$ is given by the ordinary C-bracket
\bea
\xi_{12}=2\xi^{ P}_{[1}\partial_{ P}\xi_{2]}^{ M}-\xi_{[1}^{ N}\partial^{ M}\xi_{2] N}
\eea
and $\Lambda_{12}^{AB}$ 
\be
\Lambda_{12}^{AB} = \frac{1}{2}\xi_{1}^{M}\partial_{M}{\cal P}_{\xi_{2}}^{AB} + \frac{1}{4}{\cal P}_{\xi_{1}}^{AC}{\cal P}_{\xi_{2}}^{B}{}_{C} - \left(1\leftrightarrow 2\right)\, ,
\ee
is a Lorentz parameter which ensures the closure. As in the case of the generalized Lie derivative, closure of the generalized Kosmann algebra requires imposing the strong constraint \eqref{dft-strong-constraint}.

The structure of the generalized Kosmann derivative makes manifest the geometric role of the compensating term ${\cal P}_{\xi}$. It is uniquely fixed by the simultaneous requirements of double Lorentz covariance and compatibility with generalized isometries, and acts as a field-dependent generator of local symmetry transformations associated with a generalized Killing vector. In this sense, ${\cal P}_{\xi}$ naturally plays the role of a momentum map in the doubled geometry.

Furthermore, the fact that the generalized Kosmann derivative admits a closed algebra and can be expressed entirely in terms of generalized fluxes shows that it is a fully determined, duality-covariant operator, which we therefore take as the generator of diffeomorphisms. This will serve as the starting point for the parametrization of symmetry transformations in terms of supergravity fields, which we carry out in the next subsection, where momentum maps emerge directly at the supergravity level.

\subsection{Transformation rules of heterotic supergravity fields}
\label{ss-3.2}

While generalized diffeomorphisms in DFT are usually implemented through the generalized Lie derivative, here we are interested in the effects of Lorentz-diffeomorphism covariance encoded in the generalized Kosmann derivative. We therefore replace ${\cal L}_{\xi} \rightarrow {\cal K}_{\xi}$ in the transformation rules, leading to
\be
\label{dft-kosmann-transformations}
\delta E_{M}{}^{A} = {\cal K}_{\xi}E_{M}{}^{A} + E_{M}{}^{B}\Lambda_{B}{}^{A}\, , \qquad \delta d = {\cal K}_{\xi}d\, .
\ee
As shown above, the generalized Kosmann derivative defines a well-posed operator on the doubled space, since it can be written entirely in terms of generalized fluxes. To explicitly parameterize these transformations in terms of supergravity fields, it is convenient to recall the parameterization of the generalized fluxes \eqref{gauged-fluxes}, collected in Appendix~\ref{app-fluxes}.

Using \eqref{kosmann-determined}, the action of the generalized Kosmann derivative on the generalized frame takes the explicit form
\be
{\cal K}_{\xi}E_{M}{}^{A} = {\cal L}_{\xi}E_{M}{}^{A} - \sqrt{2}\partial^{[\un{b}}\xi^{\un{a}]}E_{M\un{b}} - \sqrt{2}\partial^{[\ov{B}}\xi^{\ov{A}]}E_{M\ov{B}} - \frac{1}{\sqrt{2}}\xi^{C}F_{C}{}^{\un{ab}}E_{M\un{b}} - \frac{1}{\sqrt{2}}\xi^{C}F_{C}{}^{\ov{AB}}E_{M\ov{B}}\, ,
\ee
where we have extended the duality group from $O(D,D)$ to $O(D,D+n_g)$ in order to incorporate the heterotic gauge sector. We now proceed to derive the transformation rules of the bosonic fields of heterotic supergravity by parameterizing the transformations of the fundamental DFT fields.

We begin by considering the component $E^{\mu}{}_{\un{a}}$, whose generalized Kosmann derivative is given by
\be
{\cal K}_{\xi}E^{\mu}{}_{\un{a}} = {\cal L}_{\xi}E^{\mu}{}_{\un{a}} - \sqrt{2}\partial_{[\un{b}}\xi_{\un{a}]}E^{\mu\un{b}} - \frac{1}{\sqrt{2}}\xi^{\un{c}}F_{\un{cab}}E^{\mu\un{b}} - \frac{1}{\sqrt{2}}\xi^{\ov{c}}F_{\ov{c}\un{ab}}E^{\mu\un{b}} - \frac{1}{\sqrt{2}}\xi^{\ov{i}}F_{\ov{i}\un{ab}}E^{\mu\un{b}}\, . 
\ee
Substituting the parameterization of the generalized frame \eqref{frame-gauge-param} together with the fluxes listed in Appendix \ref{app-fluxes}, we obtain
\be
\label{kosmann-vielbein-down}
{\cal K}_{\xi}E^{\mu}{}_{\un{a}} = L_{\xi}e^{\mu}{}_{\un{a}} + \xi^{\nu}w_{\nu \un{ba}}e^{\mu \un{b}} + \nabla_{[\un{a}}\xi_{\un{b}]}e^{\mu \un{b}} + \frac{1}{2}g^{\rho\mu}e^{\nu}{}_{\un{a}}\left[L_{\xi}b_{\rho\nu} + \delta_{\lambda,\chi}b_{\rho\nu} + A_{[\rho}{}^{i}L_{\xi}A_{\nu]i} + A_{[\rho}{}^{i}\delta_{\chi}A_{\nu]i}\right]\, .
\ee
In this expression, we recognize the standard Kosmann derivative \eqref{Kosmann-derivative} acting on the inverse vielbein $e^{\mu}{}_{a}$, supplemented by additional terms involving the transformations of the Kalb-Ramond and Yang-Mills fields under diffeomorphisms and gauge transformations
\be
\delta_{\lambda,\chi}b_{\mu\nu} = 2\partial_{[\mu}\lambda_{\nu]} - \partial_{[\mu}\chi^{i}A_{\nu]i}\, , \qquad \delta_{\chi}A_{\mu}{}^{i} = \partial_{\mu}\chi^{i} + f^{i}{}_{jk}\chi^{j}A_{\mu}{}^{k}\, .
\ee

The computation for the component $E^{\mu}{}_{\ov{a}}$ proceeds analogously. One finds
\be
{\cal K}_{\xi}E^{\mu}{}_{\ov{a}} = {\cal L}_{\xi}E^{\mu}{}_{\ov{a}} - \sqrt{2}\partial_{[\ov{B}}\xi_{\ov{a}]}E^{\mu\ov{B}} - \frac{1}{\sqrt{2}}\xi^{\un{c}}F_{\un{c}\ov{aB}}E^{\mu\ov{B}} - \frac{1}{\sqrt{2}}\xi^{\ov{C}}F_{\ov{CaB}}E^{\mu\ov{B}}\, , 
\ee
which leads to
\be
\label{kosmann-vielbein-up}
{\cal K}_{\xi}E^{\mu}{}_{\ov{a}} = L_{\xi}e^{\mu}{}_{\ov{a}} + \xi^{\nu}w_{\nu \ov{ba}}e^{\mu \ov{b}} + \nabla_{[\ov{a}}\xi_{\ov{b}]}e^{\mu \ov{b}} - \frac{1}{2}g^{\rho\mu}e^{\nu}{}_{\ov{a}}\left[L_{\xi}b_{\rho\nu} + \delta_{\lambda,\chi}b_{\rho\nu} + A_{[\rho}{}^{i}L_{\xi}A_{\nu]i} + A_{[\rho}{}^{i}\delta_{\chi}A_{\nu]i}\right]\, ,
\ee

At first sight, equations \eqref{kosmann-vielbein-down} and \eqref{kosmann-vielbein-up} may appear contradictory, since both should describe the action of the generalized Kosmann derivative on the supergravity vielbein once we identify $e_{\mu}{}^{\un{a}}=e_{\mu}{}^{\ov{a}}=e_{\mu}{}^{a}$, with the latter being the supergravity vielbein. This apparent tension is solved by noting that the corresponding components of the generalized frame transform under different Lorentz groups. As a consequence, a gauge fixing is required in order to establish a precise match with supergravity. To this end, it is convenient to consider the full transformations
\bea
\delta E^{\mu}{}_{\un{a}} = {\cal K}_{\xi}E^{\mu}{}_{\un{a}} + \Lambda_{\un{ba}}E^{\mu\un{b}} & \rightarrow & \delta e^{\mu}{}_{a} = K_{\xi}e^{\mu}{}_{a} + \frac{1}{2}e^{\rho}{}_{b}e^{\nu}{}_{a}\left(\delta b_{\rho\nu} + A_{[\rho}{}^{i}\delta A_{\nu]i}\right)e^{\mu b} - \un{\Lambda}_{ba}e^{\mu b}\, \nn \\
\delta E^{\mu}{}_{\ov{a}} = {\cal K}_{\xi}E^{\mu}{}_{\ov{a}} + \Lambda_{\ov{ba}}E^{\mu\ov{b}} & \rightarrow & \delta e^{\mu}{}_{a} = K_{\xi}e^{\mu}{}_{a} - \frac{1}{2}e^{\rho}{}_{b}e^{\nu}{}_{a}\left(\delta b_{\rho\nu} + A_{[\rho}{}^{i}\delta A_{\nu]i}\right)e^{\mu b} + \ov{\Lambda}_{ba}e^{\mu b}\, \nn ,
\eea
where the components of the double Lorentz parameter are expressed in terms of standard Lorentz indices as
\be
\Lambda_{\ov{ab}}\delta^{\ov{a}}_{a}\delta^{\ov{b}}_{b} = \ov{\Lambda}_{ab}\, ,\qquad  \Lambda_{\un{ab}}\delta^{\un{a}}_{a}\delta^{\un{b}}_{b} = \un{\Lambda}_{ab}\, .
\ee

Consistency requires that both variations of the supergravity vielbein coincide. This condition imposes the following relation between the double Lorentz parameters
\be
\label{definition-of-delta-b}
\un{\Lambda}_{ab} + \ov{\Lambda}_{ab} = e^{\mu}{}_{a}e^{\nu}{}_{b}\left(L_{\xi}b_{\mu\nu} + \delta_{\lambda,\chi}b_{\mu\nu} + A_{[\mu}{}^{i}L_{\xi}A_{\nu]i} + A_{[\mu}{}^{i}\delta_{\chi}A_{\nu]i}\right) = \delta b_{ab} + \delta A_{ab}\, .
\ee
Different solutions to this equation are related by redefinitions of the standard Lorentz parameter of supergravity, $\Lambda_{ab}$. We choose
\be
\un{\Lambda}_{ab} = - \Lambda_{ab} + \frac{1}{2}\left(\delta b_{ab} + \delta A_{ab}\right)\, , \quad \ov{\Lambda}_{ab} = \Lambda_{ab} + \frac{1}{2}\left(\delta b_{ab} + \delta A_{ab}\right)\, ,
\ee
which yields the standard Lorentz transformation of the vielbein and its inverse
\be
\label{transf-vielbein}
\delta e_{\mu}{}^{a} = K_{\xi}e_{\mu}{}^{a} + \Lambda_{b}{}^{a}e_{\mu}{}^{b}\, , \qquad \delta e^{\mu}{}_{a} = K_{\xi}e^{\mu}{}_{a} + \Lambda^{b}{}_{a}e^{\mu}{}_{b}\, .
\ee

Next, we focus on the components of the generalized frame carrying explicit gauge indices. We begin by imposing the conditions $\delta E^{\mu}{}_{\ov{i}}=0$ and $\delta E^{i}{}_{\ov{i}}=\delta e^{i}{}_{\ov{i}}$. The first condition is trivial, since this component is null, while the second follows from the fact that $e^{i}{}_{\ov{i}}$ is constant. This allows us to fix the remaining components of the double Lorentz parameter as
\be
\label{mixed-gauge-fixing}
\Lambda_{\ov{ai}}\delta^{\ov{a}}_{a}\delta^{\ov{i}}_{i} = \ov{\Lambda}_{ai} = \frac{1}{\sqrt{2}}e^{\mu}{}_{a}\left(L_{\xi}A_{\mu i} + \delta_{\chi}A_{\mu i}\right)\, , \qquad \Lambda_{\ov{ij}}e^{\ov{i}}{}_{i}e^{\ov{j}}{}_{j} = \ov{\Lambda}_{ij} = 0\, .
\ee
This gauge fixing is consistent and, moreover, highlights an interesting feature of generating diffeomorphisms via the generalized Kosmann derivative rather than through the generalized Lie derivative. In particular, the vanishing of the component $\ov{\Lambda}_{ij}$ reflects the compensating transformations induced by ${\cal K}_{\xi}$. By comparison with analogous gauge fixings in heterotic setups where generalized diffeomorphisms are implemented through ${\cal L}_{\xi}$, such as in \cite{Lescano:2021guc, Baron:2024tph}, one observes that a covariant gauge transformation previously encoded in $\ov{\Lambda}_{ij}$ is now effectively incorporated into the action of ${\cal K}_{\xi}$ itself.

The transformation of the Yang–Mills field can be extracted from several components of the generalized frame, namely $E_{\mu}{}^{\ov{i}}$, $E^{i}{}_{\ov{a}}$ or $E^{i}{}_{\un{a}}$. Among these, the component $E_{\mu}{}^{\ov{i}}$ provides the most direct route and conveniently illustrates the role of the gauge fixings introduced above. Recalling that $E_{\mu}{}^{\ov{i}}=A_{\mu}{}^{i}e_{i}{}^{\ov{i}}$, its variation reads
\be
\delta E_{\mu}{}^{\ov{i}} = {\cal K}_{\xi}E_{\mu}{}^{\ov{i}} + E_{\mu}{}^{\ov{a}}\Lambda_{\ov{a}}{}^{\ov{i}} + E_{\mu}{}^{\ov{j}}\Lambda_{\ov{j}}{}^{\ov{i}}\, .
\ee
The action of the generalized Kosmann derivative on this component is given by
\be
{\cal K}_{\xi}E_{\mu}{}^{\ov{i}} = \frac{1}{2}\Big[L_{\xi}A_{\mu i} + \delta_{\chi}A_{\mu i} + C_{\rho\mu}g^{\rho\nu}\left(L_{\xi}A_{\nu i} + \delta_{\chi}A_{\nu i}\right)\Big]e_{i}{}^{\ov{i}}\, ,
\ee
and, upon implementing the gauge-fixing conditions \eqref{mixed-gauge-fixing}, these terms combine precisely so as to reproduce the standard transformation of the Yang–Mills field
\be
\label{transformation-of-A}
\delta A_{\mu}{}^{i} = L_{\xi}A_{\mu}{}^{i} + \partial_{\mu}\chi^{i} + f^{i}{}_{jk}\chi^{j}A_{\mu}{}^{k}\, .
\ee

We now turn to the components $E_{\mu}{}^{\un{a}}$ and $E_{\mu}{}^{\ov{a}}$ in order to extract the transformation of the Kalb-Ramond field. Their variations read
\bea
\delta E_{\mu}{}^{\un{a}} & = & {\cal K}_{\xi}E_{\mu}{}^{\un{a}} + E_{\mu}{}^{\un{b}}\Lambda_{\un{b}}{}^{\un{a}}\, , \\
\delta E_{\mu}{}^{\ov{a}} & = & {\cal K}_{\xi}E_{\mu}{}^{\ov{a}} + E_{\mu}{}^{\ov{b}}\Lambda_{\ov{b}}{}^{\ov{a}} + E_{\mu}{}^{\ov{i}}\Lambda_{\ov{i}}{}^{\ov{a}}\, .
\eea
At this stage, the gauge-fixing conditions \eqref{definition-of-delta-b} and \eqref{mixed-gauge-fixing} allow us to compute the transformation of $b_{\mu\nu}$ from either expression. A straightforward, albeit somewhat lengthy, computation yields
\be
\delta b_{\mu\nu} = L_{\xi}b_{\mu\nu} + 2\partial_{[\mu}\lambda_{\nu]} - \partial_{[\mu}\chi^{i}A_{\nu]i}\, ,
\ee 
which is nothing but the standard transformation of the Kalb–Ramond field under diffeomorphisms and abelian and non-abelian gauge transformations.

Having completed the parameterization of the generalized frame transformations, we can now briefly address the generalized dilaton. Although its variation is, in principle, governed by the generalized Kosmann derivative, the fact that the dilaton transforms trivially under double Lorentz symmetry implies that its transformation reduces to that generated by the generalized Lie derivative
\be
\delta d = {\cal K}_{\xi}d = {\cal L}_{\xi}d,
\ee
which, upon parameterization, reproduces the transformation of the supergravity dilaton
\be
\delta\phi = L_{\xi}\phi = \xi^{\mu}\partial_{\mu}\phi\, .
\ee

This completes the analysis of the transformations of the bosonic sector of heterotic supergravity when generalized diffeomorphisms are generated by the generalized Kosmann derivative rather than by the generalized Lie derivative. For convenience, we collect the resulting transformations
\bea
\label{summary-of-transformations}
\delta e_{\mu}{}^{a} & = & K_{\xi}e_{\mu}{}^{a} + \Lambda_{b}{}^{a}e_{\mu}{}^{b}\, , \nn \\
\delta\phi & = & L_{\xi}\phi\, , \\
\delta b_{\mu\nu} & = & L_{\xi}b_{\mu\nu} + 2\partial_{[\mu}\lambda_{\nu]} - \partial_{[\mu}\chi^{i}A_{\nu]i}\, , \nn \\
\delta A_{\mu}{}^{i} & = & L_{\xi}A_{\mu}{}^{i} + \partial_{\mu}\chi^{i} + f^{i}{}_{jk}\chi^{j}A_{\mu}{}^{k}\, . \nn
\eea

We observe that, apart from the specific deformation of the vielbein transformation required to ensure Lorentz covariance and compatibility with isometries, the generalized Kosmann derivative does not induce additional contributions to the transformations of the remaining fields. This is expected, since the abelian gauge transformations of the Kalb–Ramond field and the Yang–Mills transformations are already encoded in the generalized diffeomorphism parameter and are therefore captured by the standard generalized Lie derivative.

In what follows, we will show that, for an appropriate choice of gauge parameters, these transformations can be naturally interpreted within the momentum-map framework introduced in \cite{Elgood:2020mdx, Elgood:2020svt} for the heterotic field content.

\subsection{Compensating transformations and the momentum maps}
\label{subsection-MM}

We have seen that the complete set of transformations summarized in \eqref{summary-of-transformations} follows directly from the generalized transformations \eqref{dft-kosmann-transformations}. A first inspection reveals that the dilaton, being a scalar, already transforms covariantly under diffeomorphisms. The vielbein, on the other hand, is acted upon by the Kosmann derivative \eqref{Kosmann-derivative} rather than by the ordinary Lie derivative, leading to a transformation law that is simultaneously covariant under diffeomorphisms and local Lorentz transformations.

Despite their origin in the generalized Kosmann derivative, the transformation rules for the Kalb--Ramond and Yang--Mills fields are not yet covariant under diffeomorphisms. Compensating transformations are therefore required. As we shall show, these compensating terms arise naturally from suitable redefinitions of the generalized diffeomorphism parameter in terms of momentum maps, closely following the mechanism discussed in \cite{Elgood:2020mdx,Elgood:2020svt}. Since the dilaton already transforms appropriately, it will be omitted from the analysis in what follows.

As a warm-up, let us briefly revisit the transformation of the vielbein, making explicit the connection with momentum maps. Its covariant behavior under diffeomorphisms is ensured by the Kosmann derivative, which yields
\be
\delta_{\xi}e_{\mu}{}^{a} = K_{\xi}e_{\mu}{}^{a} = L_{\xi}e_{\mu}{}^{a} + \xi^{\nu}w_{\nu b}{}^{a}e_{\mu}{}^{b} - \nabla^{[b}\xi^{a]}e_{\mu b}\, .
\ee
As discussed in the introduction, this transformation can be rewritten as
\be
\delta_{\xi}e_{\mu}{}^{a} = L_{\xi}e_{\mu}{}^{a} + \Lambda_{\xi b}{}^{a}e_{\mu}{}^{b}\, ,
\ee
where the parameter of the compensating Lorentz transformation is
\be
\Lambda_{\xi}^{ab} = \xi^{\nu}w_{\nu}{}^{ab} - \nabla^{[a}\xi^{b]} = \xi^{\mu}w_{\mu}{}^{ab} - P_{\xi}^{ab}\, .
\ee

This defines the momentum map $P_{\xi}^{ab}$ associated with the vielbein $e_{\mu}{}^{a}$. The diffeomorphism transformation can equivalently be written in the gauge-covariant way
\be
\delta_{\xi}e_{\mu}{}^{a} = \nabla_{\mu}\xi^{a} + P_{\xi}^{ab}e_{\mu b}\, .
\ee
The interpretation of $P_{\xi}^{ab}$ as a momentum map is further supported by considering the case in which $\xi^{\mu}$ is a Killing vector. In this situation one finds
\be
\label{relationRiem}
\xi^{\nu}R_{\nu\mu}{}^{ab}
= - \nabla_{\mu}P_{\xi}^{ab}\, .
\ee

We now turn to the Yang-Mills field $A_{\mu}{}^{i}$. Before analyzing its full symmetry transformation, it is convenient to rewrite its Lie derivative in terms of the field strength $F_{\mu\nu}{}^{i}$, which is a diffeomorphism-covariant, gauge-invariant tensor. One finds
\be
L_{\xi}A_{\mu}{}^{i} = \xi^{\nu}F_{\nu\mu}{}^{i} + \partial_{\mu}\!\left(\xi^{\nu}A_{\nu}{}^{i}\right) + f^{i}{}_{jk}\left(\xi^{\nu}A_{\nu}{}^{j}\right)A_{\mu}{}^{k}\, .
\ee
The covariance under gauge transformation is restored by supplementing the diffeomorphism with a compensating transformation parametrized by $\chi_{\xi}^{i}$, leading to the Lie-Maxwell derivative \cite{Elgood:2020svt}
\be
\label{difeo-covariante-A}
\delta_{\xi}A_{\mu}{}^{i} = \xi^{\nu}F_{\nu\mu}{}^{i} + \partial_{\mu}\left(\xi^{\nu}A_{\nu}{}^{i} + \chi_{\xi}^{i}\right) + f^{i}{}_{jk}\left(\xi^{\nu}A_{\nu}{}^{j} + \chi_{\xi}^{j}\right)A_{\mu}{}^{k}\, .
\ee

Beyond covariance, we require invariance under isometries, namely $\delta_{\xi}A_{\mu}{}^{i}=0$ when $\xi^{\mu}$ is a Killing vector field. As shown in \cite{Elgood:2020mdx}, invariance of the field strength guarantees the existence of a momentum map $P_{\xi}^{i}$, satisfying
\be
\xi^{\nu}F_{\nu\mu}{}^{i} = - \partial_{\mu}P_{\xi}^{i} - f^{i}{}_{jk}P_{\xi}^{j}A_{\mu}{}^{k}\, .
\ee
Comparing with \eqref{difeo-covariante-A}, the diffeomorphism transformation can be written as the sum of a Lie derivative and a compensating gauge transformation,
\be
\delta_{\xi}A_{\mu}{}^{i} = L_{\xi}A_{\mu}{}^{i} + \partial_{\mu}\chi_{\xi}^{i} + f^{i}{}_{jk}\chi_{\xi}^{j}A_{\mu}{}^{k}\, ,
\ee
which is compatible with invariance under isometries provided
\be
\label{gauge-compensating-parameter}
\chi_{\xi}^{i} = P_{\xi}^{i} - \xi^{\mu}A_{\mu}{}^{i}\, .
\ee

Remarkably, the same result follows directly from DFT through a simple redefinition of the component $\xi^{i}$ of the generalized diffeomorphism parameter,
\be
\label{redef-gauge-param}
\xi^{i} = \chi^{i} \longrightarrow \chi^{i} + P_{\xi}^{i} - \xi^{\mu}A_{\mu}{}^{i}\, ,
\ee
from which the compensating terms arise naturally.

We still need to analyze the Kalb-Ramond field. As before, we begin by rewriting its Lie derivative in terms of the gauge-invariant curvature $H_{\mu\nu\rho}$,
\be
L_{\xi}b_{\mu\nu} = \xi^{\rho}H_{\rho\mu\nu} + \partial_{\mu}\left(\xi^{\rho}b_{\rho\nu}\right) - \partial_{\nu}\left(\xi^{\rho}b_{\rho\mu}\right) + 3\xi^{\rho}A_{[\mu}{}^{i}\partial_{\nu}A_{\rho]i} - \xi^{\rho}f_{ijk}A_{\mu}{}^{i}A_{\nu}{}^{j}A_{\rho}{}^{k}\, .
\ee
In addition to diffeomorphisms, the Kalb--Ramond field is subject to both abelian and non-abelian gauge symmetries. The compensating non-abelian transformation is fixed by \eqref{gauge-compensating-parameter}, while the remaining contribution arises from the abelian gauge transformation and is parameterized by $\lambda_{\xi \mu}$, yielding
\be
\label{difeo-covariante-B}
\delta_{\xi}b_{\mu\nu}
= \xi^{\rho}H_{\rho\mu\nu} + P_{\xi}^{i}F_{\mu\nu i} + 2\partial_{[\mu}\left(\xi^{\rho}b_{|\rho|\nu]} + \lambda_{\xi\nu]} - \frac{1}{2}P_{\xi}^{i}A_{\nu]i}\right) - A_{[\mu}{}^{i}\delta_{\xi}A_{\nu]i}\, .
\ee

Invariance of the three-form $H_{\mu\nu\rho}$ under isometries implies the existence of a momentum map $P_{\xi\mu}$, satisfying
\be
\xi^{\rho}H_{\rho\mu\nu} = - 2\partial_{[\mu}P_{\xi\nu]} - P_{\xi}^{i}F_{\mu\nu i}\, .
\ee
Requiring invariance under isometries, the transformation of the Kalb-Ramond field can be written as
\be
\delta_{\xi}b_{\mu\nu} = L_{\xi}b_{\mu\nu} + 2\partial_{[\mu}\lambda_{\xi\nu]} - \partial_{[\mu}\chi_{\xi}^{i}A_{\nu]i}\, .
\ee
with
\be
\label{abelian-gauge-compensating-parameter}
\lambda_{\xi\mu} = P_{\xi\mu} + \frac{1}{2}P_{\xi}^{i}A_{\mu i} - \xi^{\nu}b_{\nu\mu}\, .
\ee

As in the Yang-Mills case, this transformation follows directly from DFT through a redefinition of the component $\xi_{\mu}$ of the generalized diffeomorphism parameter,
\be
\label{redef-abelian-gauge-param}
\xi_{\mu} = \lambda_{\mu} \longrightarrow \lambda_{\mu} + P_{\xi\mu} + \frac{1}{2}P_{\xi}^{i}A_{\mu i} - \xi^{\nu}b_{\nu\mu}\, .
\ee

Up to this point, our analysis closely follows \cite{Elgood:2020svt,Elgood:2020mdx}, with the important difference that we consider Yang--Mills fields rather than a Maxwell field, thereby incorporating non-abelian contributions in a natural way.

We conclude this subsection by emphasizing its main result: all covariant diffeomorphism transformations in heterotic supergravity can be recovered from DFT by implementing two simple ingredients:
\begin{enumerate}
\item Replacing the generalized Lie derivative with the generalized Kosmann derivative, ensuring Lorentz covariance and invariance of the vielbein through the momentum map \(P_{\xi}^{ab}\).
\item Redefining the generalized diffeomorphism parameter as
\be
\xi^{M} = \left(\xi^{\mu}, \ \ \lambda_{\mu} + P_{\xi\mu} + \frac{1}{2}P_{\xi}^{i}A_{\mu i} - \xi^{\nu}b_{\nu\mu}, \ \ \chi^{i} + P_{\xi}^{i} - \xi^{\mu}A_{\mu}{}^{i}\right)\, ,
\ee
which yields manifestly covariant transformations for $b_{\mu\nu}$ and $A_{\mu}{}^{i}$.
\end{enumerate}
This result highlights how the geometric structure of the doubled space admits natural deformations that preserve covariance and isometries in the presence of local symmetries, which are crucial in the construction of Noether currents and conserved charges.

\subsection{Geometrical aspects of \texorpdfstring{\ensuremath{\mathcal{P}_{\xi}^{AB}} vs \ensuremath{P_{\xi}^{ab}}} {P_xi_AB vs P_xi_ab}}

It is well-known that curvature tensors constructed from the generalized connection in DFT exhibit a subtle geometric structure. In particular, a direct extension of the conventional Riemann tensor
\be
R_{MNPQ} = 2\partial_{[M}\Gamma_{N]PQ} + 2\Gamma_{[M|R|Q}\Gamma_{N]P}{}^{R}\, ,
\ee
fails to transform covariantly under generalized diffeomorphisms. Unlike ordinary Riemannian geometry, this lack of covariance cannot be cured by imposing a torsionless condition. Covariance is instead restored by considering the combination
\be
\label{generalized-riemann}
{\cal R}_{MNPQ}
= R_{MNPQ} + R_{PQMN} + \Gamma_{RMN}\Gamma^{R}{}_{PQ}\, ,
\ee
which defines the generalized Riemann tensor of DFT. Although ${\cal R}_{MNPQ}$ transforms covariantly under generalized diffeomorphisms, it is not completely determined by the generalized metric or frame \cite{Hohm:2011si}, reflecting the presence of undetermined components of the DFT connection.\footnote{Despite this ambiguity, suitable projections of ${\cal R}_{MNPQ}$ give rise to well-defined generalized Ricci tensors and a generalized Ricci scalar, which encode the dynamics of the theory.}

One may ask whether a relation analogous to \eqref{relationRiem} holds in DFT, that is
\bea
\nabla_{M} {\cal P}_{\xi}^{A B} \stackrel{?}{=} - \xi^{N} {\cal R}_{N M P Q} E^{P A} E^{Q B} + \tilde \Delta_{M}^{A B} = - \xi^{N} {R}_{N M P Q} E^{P A} E^{Q B} + \Delta_{M}^{A B} \, ,
\eea
where $\tilde \Delta_{M}^{A B}$ denotes a quantity covariant under generalized diffeomorphisms, while $\Delta_{M}^{A B}$ is non-covariant. To analyze this relation, it is convenient to consider the commutator of two covariant derivatives
\bea
[\nabla_{M}, \nabla_{N}] \xi_{P} = - R_{M N P}{}^{Q} \xi_{Q} - 2 \Gamma_{[M N]}{}^{Q} \nabla_{Q} \xi_{P} \, ,
\label{commutator}
\eea
and to introduce the notation ${\cal P}_{\xi M N}= \nabla_{[M} \xi_{N]}$. Using \eqref{commutator}, one finds
\bea
\nabla_{M} {\cal P}_{\xi N P} = \nabla_{[N|} \nabla_{M} \xi_{|P]} - R_{M [N P]}{}^{Q} \xi_{Q} + \Gamma^{Q}{}_{M [N|} \nabla_{Q} \xi_{|P]} \, .
\eea
Upon contraction with generalized frames, this expression becomes
\bea
\nabla_{M} {\cal P}_{\xi A B} = \nabla_{[A|} \nabla_{M} \xi_{|B]} -  E^{N}{}_{[A} E^{P}{}_{B]} R_{M N P}{}^{Q} \xi_{Q} + \Gamma^{Q}{}_{M N} E^{N}{}_{[A}  \nabla_{Q} \xi_{|B]} \, .
\label{MomentumrelationDFT}
\eea

From \eqref{MomentumrelationDFT}, both $\tilde \Delta_{M}^{A B}$ and $\Delta_{M}^{A B}$ can be identified explicitly. In contrast to the Riemannian case, these extra contributions do not vanish. As a consequence, the covariant derivative of ${\cal P}_{\xi}^{AB}$ does not reduce to a simple contraction between the generalized curvature and the generating vector field.

This result shows that the geometrical interpretation of ${\cal P}_{\xi}^{AB}$ differs fundamentally from that of the ordinary momentum map $P_{\xi}^{ab}$. In general relativity, the Killing condition together with the Bianchi identity ensures the cancellation of all additional terms, leading directly to \eqref{relationRiem}. In DFT, however, the doubled geometry introduces extra contributions that are not constrained by the generalized Killing or Bianchi identities. These terms, encoded in $\tilde \Delta_{M}^{AB}$ and $\Delta_{M}^{AB}$, obstruct a direct momentum-map interpretation purely in terms of curvature.

More generally, the presence of these non-vanishing contributions reflects the richer geometric structure underlying DFT. In particular, ${\cal P}_{\xi}^{AB}$ encodes information not only about generalized isometries, but also about the embedding of physical spacetime into the doubled phase space \cite{Lescano:2020smc}.

\section{Noether charges, local symmetries and black hole thermodynamics}

In any diffeomorphism-invariant theory of gravity, black hole entropy can be identified, via the first law of black hole mechanics, with a Noether charge associated with diffeomorphism invariance \cite{Wald:1993nt, Iyer:1994ys}. When applied to general relativity, this construction reproduces the Bekenstein-Hawking entropy and provides a systematic framework to generalize the notion of black hole entropy to theories with higher-derivative interactions.

This picture is modified in the presence of local symmetries, for which the standard Noether-charge construction is no longer sufficient. In such cases, additional contributions arise in the first law, typically interpreted as work terms associated with local Lorentz or gauge transformations. Suitable extensions of the first law are therefore required in order to consistently incorporate these symmetries, as discussed in several works \cite{Jacobson:2015uqa, Prabhu:2015vua, Edelstein:2018ewc, Edelstein:2019wzg, Aneesh:2020fcr, Elgood:2020mdx, Elgood:2020svt, Ortin:2022uxa, Ballesteros:2023iqb, Ballesteros:2023muf, Ballesteros:2024prz, Ballesteros:2025wvs}. In particular, \cite{Jacobson:2015uqa} showed that this can be achieved by replacing the Lie derivative with the Kosmann derivative. Although momentum maps are not explicitly discussed there, their role is intrinsic to this construction, as clarified in \cite{Elgood:2020mdx}.

In DFT, black hole thermodynamics was first studied in \cite{Arvanitakis:2016zes}, where generalized currents and charges \cite{Park:2015bza, Blair:2015eba, Naseer:2015fba} were employed to extend the first law and to derive duality-invariant expressions for mass and entropy. This analysis was performed within the generalized metric formulation of DFT and therefore does not allow for a systematic treatment of local symmetries. It is thus natural to ask how this approach should be modified in order to consistently incorporate local symmetries, in light of the results of Section \ref{section-3}. We begin by briefly reviewing some key aspects of the construction presented in \cite{Arvanitakis:2016zes} before discussing its extension.

The variation of the DFT action under a generalized diffeomorphism parameterized by $\xi^{M}$ yields a boundary term, leading to an on-shell conserved current
\be
J^{M} = e^{-2d}\left(\Theta^{M}[\Phi;{\cal L}_{\xi}\Phi] - \xi^{M}{\cal R}\right)\, , \qquad \partial_{M}J^{M} = 0\, \ \ \text{(on-shell)}\, , 
\ee
where $\Phi$ collectively denotes the DFT fields, ${\cal R}$ is the generalized curvature defined in \eqref{generalized-Ricci-H}, and $\Theta^{M}$ is the symplectic potential
\bea
\label{symplectic-potential}
\Theta^{M} & = & \delta H^{KL}\left(\frac{1}{4}H^{MN}\partial_{N}H_{KL} - \frac{1}{2}H^{MN}\partial_{K}H_{NL} - \frac{1}{2}H^{N}{}_{K}\partial_{N}H^{M}{}_{L}\right)\, \nn \\
& & + 4H^{MN}\partial_{N}\delta d - \partial_{N}\delta H^{MN} + 2\delta H^{MN}\partial_{N}d\, ,
\eea
obtained as the boundary term arising from the variation of the action. On-shell conservation of the current implies the existence of an antisymmetric tensor density such that
\be
J^{M} = \partial_{N}J^{MN}\, ,
\ee
which, when integrated over a codimension 2 surface at infinity, contributes to the Noether charge associated to $\xi^{M}$,
\be
\label{charge}
Q_{\xi} = \int_{\partial C(\infty)}\left(J^{MN} - 2e^{-2d}B^{[M}\xi^{N]}\right)\varepsilon_{MN} = \int_{\partial C(\infty)}Q^{MN}\varepsilon_{MN}\, ,
\ee
where $B^{M}$ is a boundary vector ensuring finiteness under Dirichlet boundary conditions. The first law of black hole thermodynamics then follows from a variational identity involving this Noether charge.

In the presence of local symmetries, however, the Noether procedure requires conserved currents that are invariant under the full symmetry group of the theory. This, in turn, demands a notion of symmetry generator that acts covariantly on the field content. In double geometry, this role is played by the generalized Kosmann derivative \eqref{Kosmann-schem}, in close analogy with the role of the Kosmann derivative in conventional gravity and in agreement with the analysis of \cite{Jacobson:2015uqa}.

Following \cite{Jacobson:2015uqa}, the appropriate strategy consists in replacing
\be
{\cal L}_{\xi}\Phi \longrightarrow {\cal K}_{\xi}\Phi^{e}\, ,
\ee
in the construction of the conserved current and the associated Noether charge, where the fields $\Phi^{e}$ now carry internal symmetry indices labeled by $e$. Accordingly, the symplectic potential must be derived from the action written in terms of generalized fluxes \eqref{dft-action}, using Kosmann rather than Lie variations. This leads to a modified conserved current of the form
\be
J^{M} = e^{-2d}\left(\Theta^{M}[\Phi^{e};{\cal K}_{\xi}\Phi^{e}] - \xi^{M}{\cal R}\right)\, .
\ee
The charge density appearing in \eqref{charge} must be adjusted accordingly in order to ensure both finiteness and gauge invariance.

This framework allows DFT to incorporate local gauge symmetries in a systematic and duality-covariant manner. As shown by the emergence of momentum maps from DFT, a duality-covariant description of black hole thermodynamics in the presence of gauge symmetries naturally follows. This opens the door to the treatment of physically relevant scenarios involving electric and magnetic potentials \cite{Elgood:2020svt, Ortin:2022uxa}. In particular, the inclusion of magnetic charge contributions addresses a limitation of the analysis presented in \cite{Arvanitakis:2016zes}. A more detailed and comprehensive investigation along these lines will be presented in a forthcoming work \cite{Ballesteros:BH-Tduality-WIP}.

\subsection{Higher-derivative corrections}
\label{hdc}
One of the main strengths of the flux formulation of DFT is its ability to consistently incorporate higher-derivative deformations, encoding the $\ap$-corrections of bosonic and heterotic supergravity,\footnote{See \cite{Lescano:2021lup} for a pedagogical introduction to this topic.} as well as more general theories beyond string theory, such as the HSZ theory \cite{Hohm:2013jaa}. In particular, the complete set of four-derivative corrections compatible with T-duality was classified in \cite{Marques:2015vua}, where it was shown that the generalized metric formulation of DFT is insufficient to describe these effects in a manifestly duality-invariant manner, while the flux formulation provides a natural and systematic framework. A key development was the observation that higher-derivative corrections can be reformulated as a two-derivative theory by enlarging the duality group from $O(D,D)$ to $O(D+k,D+k)$, via the so-called generalized Bergshoeff-de Roo (gBdR) identification \cite{Baron:2018lve, Baron:2020xel}, providing a natural setting to analyze $\ap$-corrections to black hole thermodynamics. 

Within this enlarged duality group, the generalized Kosmann derivative can be defined in complete analogy with the two-derivative theory as
\be
\hat{{\cal K}}_{\hat{\xi}}{\cal E}_{{\cal M}}{}^{{\cal A}} = \hat{{\cal L}}_{\hat{\xi}}{\cal E}_{{\cal M}}{}^{{\cal A}} - \sqrt{2}\partial^{[\un{\cal B}}\hat{\xi}^{\un{\cal A}]}{\cal E}_{{\cal M}\un{{\cal B}}} - \sqrt{2}\partial^{[\ov{\cal B}}\hat{\xi}^{\ov{\cal A}]}{\cal E}_{{\cal M}\ov{{\cal B}}} - \frac{1}{\sqrt{2}}\hat{\xi}^{\cal C}{\cal F}_{\cal C}{}^{\un{{\cal A}{\cal B}}}{\cal E}_{{\cal M}\un{{\cal B}}} - \frac{1}{\sqrt{2}}\hat{\xi}^{\cal C}{\cal F}_{\cal C}{}^{\ov{{\cal A}{\cal B}}}{\cal E}_{{\cal M}\ov{{\cal B}}}\, ,
\ee
where calligraphic symbols distinguish $O(D+k,D+k)$ quantities from their $O(D,D)$ counterparts. The associated symplectic potential retains the same functional form as in the lowest-order flux formulation, now expressed in terms of the extended generalized metric and dilaton. Upon implementing the gBdR identification, and performing the replacement $\hat{\cal L}_{\hat{\xi}}\rightarrow\hat{\cal K}_{\hat{\xi}}$ in the variational analysis, this construction allows one to systematically extract higher-derivative corrections to the symplectic potential 
\be
\hat{\Theta}^{\cal M} = \Theta^{M} + {\cal O}(\ap)\, ,
\ee
and consequently to the conserved currents and charges
\be
\hat{\cal J}^{\cal M} = J^{M} + {\cal O}(\ap)\, , \qquad \hat{\cal Q}_{\hat{\xi}} = Q_{\xi} + {\cal O}(\ap)\, .
\ee
This approach might provide a duality-covariant extension of the first law of black hole mechanics including higher-derivative corrections.

\section{Concluding Remarks}

In this work we have focus on the construction of the generalized Kosmann derivative within the flux formulation of Double Field Theory (DFT). Our main result is the identification and explicit construction of the compensating transformation required to ensure double Lorentz covariance and compatibility with generalized isometries, while rendering the generalized Kosmann derivative fully determined. In this way, our construction generalizes the semi-covariant approach of \cite{Angus:2018mep} to a framework where the geometry is encoded in generalized fluxes.

The compensating term plays the role of a generalized momentum map and is given explicitly by ${\cal P}_{\xi}^{AB}$ in \eqref{generalized-momentum-map}. Its inclusion ensures the cancellation of all undetermined components of the generalized spin connection that arise when imposing double Lorentz covariance on the generalized Lie derivative, yielding a fully covariant and well-defined operator acting on multiplets of the duality group, as shown in \eqref{kosmann-determined}. We have further shown that the algebra of generalized Kosmann transformations closes upon imposing the strong constraint, in full agreement with the structure of doubled geometry and its role in ensuring the consistency of generalized diffeomorphisms.

We have also investigated the geometrical role of the compensating term, interpreting it as a generalized momentum map in doubled geometry. In particular, we studied the action of the generalized covariant derivative on this object and found additional contributions when compared to its defining property in Riemannian geometry. These extra terms are required in order to ensure manifest T-duality invariance and reflect the richer interplay between isometries and curvature in the doubled phase space. A more detailed analysis of this relation could provide further insight into the geometric structure underlying DFT.

By explicitly parameterizing the generalized Kosmann derivative in terms of the bosonic field content of heterotic supergravity, we have shown how covariant diffeomorphisms and gauge transformations emerge from a duality-covariant framework. In particular, covariance and isometry invariance of the vielbein are encoded through the appearance of the standard Kosmann derivative and the associated momentum map $P_{\xi}^{ab}$, while for the remaining fields these properties arise through a redefinition of the generalized diffeomorphism parameter. This makes manifest that double geometry provides a unified and intrinsic description of covariant diffeomorphisms in the presence of local symmetries at the supergravity level, rather than requiring additional structures imported from the lower-dimensional theory.

This observation has direct implications for the construction of conserved currents and Noether charges in doubled geometry. In particular, the existence of a fully covariant symmetry generator acting on generalized frames provides the appropriate starting point for defining conserved quantities when internal symmetries are present. This is especially relevant for applications to black hole thermodynamics, where internal symmetries contribute to the first law, typically through thermodynamic potentials. In the language of momentum maps, for instance, these are associated with the electric \cite{Elgood:2020svt,Prabhu:2015vua} and magnetic potentials \cite{Ortin:2022uxa}, as well as with the surface gravity and hence the black hole temperature \cite{Prabhu:2015vua}, once evaluated at the bifurcation surface. Our results indicate that a duality-covariant formulation of these structures is naturally accommodated within DFT once local symmetries are properly incorporated, extending previous studies of black hole thermodynamics in doubled geometry \cite{Arvanitakis:2016zes}.

Since our analysis is performed within the flux formulation of DFT, it is compatible with the generalized Bergshoeff-de Roo identification \cite{Baron:2018lve, Baron:2020xel}, allowing $\alpha'$ corrections to be incorporated while preserving manifest duality covariance. This opens the possibility of systematically studying higher-derivative contributions to conserved currents, Noether charges, and black hole thermodynamics in string-inspired gravitational theories. While a complete treatment of higher-order corrections requires addressing additional subtleties, such as those associated with the generalized Green-Schwarz mechanism \cite{Marques:2015vua}, the framework developed here provides the necessary geometric and symmetry-theoretic foundation for such investigations.


\subsection*{Acknowledgements}

The work of R.B. is supported by the Vicerrectoría de Investigación y Doctorados of Universidad San Sebastián through the postdoctoral project USS-FIN-25-PDOC-05. E.L. is supported by the SONATA BIS grant 2021/42/E/ST2/00304 from the National Science Centre (NCN), Poland. J.A.R. acknowledge financial support from UADE. 

\newpage

\appendix

\section{Supergravity conventions}

For a generic tensor field carrying space-time, Lorentz, and Yang--Mills indices, the covariant derivative is defined as
\bea
\nabla_{\rho}T_{\mu a}^{\nu i} & = & \partial_{\rho}T_{\mu a}^{\nu i} - \Gamma_{\rho\mu}^{\lambda}T_{\lambda a}^{\nu i} + \Gamma_{\rho\lambda}^{\nu}T_{\mu a}^{\lambda i} - w_{\rho a}{}^{b}T_{\mu b}^{\nu i} - f^{i}{}_{jk}A_{\rho}{}^{j}T_{\mu a}^{\nu k}\, ,
\eea
where $\Gamma_{\mu\nu}^{\rho}$ is the affine connection, $w_{\mu ab}$ is the spin connection, and $A_{\mu}{}^{i}$ is the Yang-Mills gauge field. Infinitesimal Lorentz transformations act on Lorentz vectors through an antisymmetric parameter $\Lambda_{ab}$ as
\be
\delta_{\Lambda}T_{a} = T_{b}\Lambda^{b}{}_{a}\, .
\ee

Assuming vanishing torsion and metric (vielbein) compatibility, the affine (spin) connection reduce to 
\be
\Gamma_{\mu\nu}^{\rho} = \frac{1}{2}g^{\rho\lambda}\left(\partial_{\mu}g_{\nu\lambda} + \partial_{\nu}g_{\mu\lambda} - \partial_{\lambda}g_{\mu\nu}\right)\, ,
\ee
and
\be
\label{spin-connection}
w_{\mu bc} = e_{\mu}{}^{a}w_{abc} = e_{\mu}{}^{a}\left(e^{\nu}{}_{[a}\partial_{b]}e_{\nu c} - e^{\nu}{}_{[a}\partial_{c]}e_{\nu b} - e^{\nu}{}_{[b}\partial_{c]}e_{\nu a}\right)\, .
\ee

The Yang-Mills and Kalb-Ramond field strengths are defined as
\be
\label{F-ym}
F_{\mu\nu}{}^{i} = 2\partial_{[\mu}A_{\nu]}{}^{i} - f^{i}{}_{jk}A_{\mu}{}^{j}A_{\nu}{}^{k}\, , \qquad
H_{\mu\nu\rho} = 3\left(\partial_{[\mu}b_{\nu\rho]} - A_{[\mu}{}^{i}\partial_{\nu}A_{\rho]i} + \frac{1}{3}f_{ijk}A_{\mu}{}^{i}A_{\nu}{}^{j}A_{\rho}{}^{k}\right)\, .
\ee

\section{Parameterization of the generalized fluxes}
\label{app-fluxes}

In this appendix we collect the explicit parameterization of the generalized fluxes \eqref{gauged-fluxes}, as obtained from the parameterization of the generalized frame given in \eqref{frame-gauge-param}
\bea
\label{param-fluxes}
\begin{alignedat}{5} 
F_{\un{abc}} & \doteq - 3\omega_{[abc]} - \frac{1}{2}H_{abc}\, , & \qquad \qquad & F_{\ov{abc}} \doteq 3\omega_{[abc]} - \frac{1}{2}H_{abc}\, , \\
F_{\ov{a}\un{bc}} & \doteq - \omega_{abc} - \frac{1}{2}H_{abc}\, , & \qquad \qquad & F_{\un{a}\ov{bc}} \doteq \omega_{abc} - \frac{1}{2}H_{abc}\, , \\
F_{\un{ab}}{}^{\ov{i}} & \doteq - \frac{1}{\sqrt{2}}e^{\mu}{}_{a}e^{\nu}{}_{b}F_{\mu\nu}{}^{i}\, , & \qquad \qquad & F_{\un{a}\ov{b}}{}^{\ov{i}} \doteq - \frac{1}{\sqrt{2}}e^{\mu}{}_{a}e^{\nu}{}_{b}F_{\mu\nu}{}^{i}\, , \\
F_{\ov{ab}}{}^{\ov{i}} & \doteq - \frac{1}{\sqrt{2}}e^{\mu}{}_{a}e^{\nu}{}_{b}F_{\mu\nu}{}^{i}\, , & \qquad \qquad & F_{\un{a}\ov{ij}} \doteq - e^{\mu}{}_{a}A_{\mu}{}^{k}f_{ijk}\, , \\
F_{\ov{aij}} & \doteq - e^{\mu}{}_{a}A_{\mu}{}^{k}f_{ijk}\, , & \qquad \qquad & F_{\ov{ijk}} \doteq \sqrt{2}f_{ijk}\, ,
\end{alignedat}
\eea
where $\doteq$ represents parametrization plus gauge fixing.

\bibliographystyle{JHEP}
\bibliography{references}

\end{document}